\documentclass[aps,prd,twocolumn,superscriptaddress,floatfix, nofootinbib, groupedaddress]{revtex4}  
\usepackage{cancel}
\usepackage{graphicx,units}
\usepackage{amssymb}
\usepackage{graphicx}
\usepackage{dcolumn}
\usepackage{float}
\usepackage{bm}
\usepackage{amsmath}
\usepackage{cancel}
\usepackage{graphicx,units}
\usepackage{amssymb}
\usepackage{graphicx}
\usepackage{dcolumn}
\usepackage{float}
\usepackage{bm}
\usepackage{amsmath}

\usepackage{soul,xcolor}
\usepackage[normalem]{ulem}
\newcommand{\bw}[1]{\textcolor{red}{[BW: #1]}}
\newcommand{\aht}[1]{{#1}}

\usepackage{appendix}

\begin{document}

\preprint{APS/123-QED}

\title{The First-Order Velocity Memory Effect from Compact Binary Coalescing Sources}
\author{Atul K. Divakarla}
 \email{akd703@ufl.edu}
 \affiliation{%
 Department of Physics, University of Florida, 2001 Museum Road, Gainesville, FL 32611-8440, USA \\
 }
 
\author{Bernard F. Whiting}
 \email{bwhiting@ufl.edu}
 \affiliation{%
 Department of Physics, University of Florida, 2001 Museum Road, Gainesville, FL 32611-8440, USA \\
 }

\begin{abstract}
It has long been known that gravitational waves from compact binary coalescing sources are responsible for a first-order displacement memory effect experienced by a pair of freely falling test masses.
This constant displacement is sourced from the non-vanishing final gravitational-wave strain present in the wave's after-zone, often referred to as the non-linear memory effect, and is of the same order of magnitude as the strain from the outgoing quadrupole radiation.
Hence, this prediction of general relativity is verifiable experimentally by measurement of 
the final relative separation between test masses that comprise gravitational-wave detectors.
In a separate context, independent calculations have demonstrated
that exact, sandwich, plane wave spacetimes exhibit a velocity memory effect: a non-zero relative velocity, gained by a pair of test masses in free fall, after the passage of a gravitational wave.
In this paper, we find that in addition to the known constant displacement memory effect test masses experience, a velocity memory effect at leading order arises due to the non-linear nature of gravitational waves from compact binary sources. 
We discuss the magnitude of the first-order velocity memory effect in the context of observing gravitational-wave radiation from super massive binary black hole mergers in LISA.
\end{abstract}

\maketitle

\section{\label{sec:level1}Introduction}

Since 2015, gravitational waves from dozens of compact binary coalescence (CBC) sources have been detected by the LIGO and Virgo experiments  ~\cite{LIGOScientific:2018mvr,Abbott:2020niy}. 
CBC sources are astrophysical pairs of compact objects, such as black holes, white dwarfs, or neutron stars, that emit gravitational-wave radiation as they merge together from orbiting under the pull of each other's gravity. 
Observation of such high energy astrophysical events now allows physicists to test the rich predictions of the theory of general relativity
using gravitational-wave strain data ~\cite{Abbott:2020jks, TheLIGOScientific:2016src, LIGOScientific:2019fpa, PhysRevLett.123.011102, Lasky:2016knh, Divakarla:2019zjj, Ebersold:2020zah, Hubner:2019sly, Khera:2020mcz}.


One of the curious predictions of the full non-linear theory of general relativity is memory effects -- physically observable phenomenon that leave the final state of gravitational-wave detectors ever so slightly altered with respect to their initial undisturbed state. 
Memory effects are of interest to those who study general relativity  because they are found to have direct ties to asymptotic symmetry and soft graviton theorems ~\cite{Strominger:2017zoo, Strominger:2014pwa, Zhang:2017geq}, as well as the black hole information paradox ~\cite{PhysRevLett.116.231301}.
Calculations of memory effects using asymptotically flat spacetimes, for example using the Bondi–Metzner–Sachs formulation \cite{osti_4799323, PhysRev.128.2851}, have been carried out in \cite{Strominger:2014pwa, Mitman:2020pbt, Tahura:2020vsa, Tolish:2014bka, Seraj:2021qja}.
Tests of gravitational-wave memory effects from astrophysical sources have been conducted using gravitational-wave strain data from the LIGO and Virgo detectors in \cite{Hubner:2019sly,Lasky:2016knh, Khera:2020mcz, Ebersold:2020zah}, from the Parkes Pulsar Timing Array \cite{Wang:2014zls}, and from NANOGrav \cite{Aggarwal:2019ypr, Arzoumanian:2015cxr}.
Signal-to-noise ratios of the memory effect from a population of super massive binary black hole mergers have been computed with respect to the LISA detector's strain spectral sensitivity in \cite{Islo:2019qht, Johnson:2018xly}.
In addition, memory effects are not unique to gravitational-wave radiation, since electromagnetic pulses can also produce memory effects \cite{Susskind:2015hpa, Bieri:2013hqa, Pasterski:2015zua, Mao:2017axa}, as can even neutrino bursts \cite{Bieri:2013gwa, Li:2017mfz}.

Memory effects can be thought of as belonging to two categories \cite{Bieri:2013ada}: null memory arises when radiation or mass-less particles escape from a system to null infinity \footnote{
Null memory which arises from non-gravitational sources is sometimes referred to as {\it linear} memory \cite{PhysRevD.46.4304}, since the effect is linear in the stress tensor of the matter fields.  We reserve the term {\it non-linear} memory \cite{PhysRevLett.67.1486} for the effects of a flux of gravitational waves to null infinity, since the flux is quadratic in (derivatives of) the strain from a gravitational-wave source.
},
and ordinary memory arises when there is a final recoil of the system relative to its initial center of mass frame 
\footnote{
\cite{Wiseman:1991ss, Favata:2009ii, PhysRevD.46.4304, PhysRevLett.67.1486} have sometimes referred to this as linear memory, but we do not choose that language.
}. 
In this paper, in so far as we follow \cite{Wiseman:1991ss} and \cite{PhysRevD.45.520}, we focus on the null memory contribution sourced from the gravitational-wave energy flux (non-linear memory) in CBC sources, although we also refer to sandwich plane gravitational waves for demonstrative purposes.

Depending upon the circumstances of the gravitational-wave detector, memory effects may manifest as a relative change in displacement or as a relative change in velocity between a pair of freely falling test masses. 
In particular, in certain interesting detection situations, asymptotic memory effects from physical sources induce first-order displacement or velocity memory effects in remotely situated detectors \cite{Zeldovich:1974gvh, PhysRevLett.67.1486, Braginsky:1986ia, Grishchuk:1989qa}.

The displacement memory effect is the phenomenon by which test masses in free fall, and initially at rest, suffer a permanent relative displacement after the passage of a gravitational wave.
One of the first examples of displacement memory was in the case of astrophysical events that emit ordinary memory, such as a flyby event between two compact binary objects \cite{Zeldovich:1974gvh}. 
It was later shown by \cite{PhysRevLett.67.1486} that the non-linear memory from compact binary sources also induces a first-order relative displacement in test masses far away.

The velocity memory effect presents itself as a non-zero difference in the relative velocity between two neighboring geodesics, after a gravitational-wave passes through their space.
To calculate the magnitude of this relative velocity, we solve the equations of motion for test particles under general initial conditions, in a particularly useful coordinate system, and study the relative motion between a pair of test masses during the gravitational wave's after-zone.

Ground-based detectors, such as LIGO and Virgo, are composed of a network of test masses that are suspended by pendulums.
These test masses effectively respond freely to the strain induced by the passage of gravitational waves travelling perpendicularly to the plane of the detector, and move predominantly along the directions of the detector arms. 
This motion is only partially free, as it is constrained to move on an arc via its point of suspension.
In principle, a velocity memory effect in these ground-based detectors would be limited by the suspension's constraint, so to fully consider the effects one would need to examine a coupled system.
In practice, controls at work within the functioning detector will counteract any velocity memory effect long before the constraining pendulum effect does so.

By contrast, LISA ~\cite{2017arXiv170200786A} is a future space-based gravitational-wave detector comprised of three freely falling test masses, housed in individual spacecrafts, forming an equilateral triangle with arm-length $L_0 = 2.5 \times 10^{9}$ meters.
LISA will observe low-frequency gravitational waves, of which super massive binary black hole mergers are the most promising sources \cite{Volonteri_2010}, for a maximum observation duration of ten years.

In the case where test masses are not initially at rest with respect to each other, such as would be expected to occur with the LISA detector \cite{2017arXiv170200786A}, we find that, as a direct generalization of the work of Christodoulou \cite{PhysRevLett.67.1486}, a velocity memory effect that is $\mathcal{O}(h)$ also occurs for gravitational-wave radiation from compact binary sources. 
This first-order velocity memory effect is sourced by the non-linear memory effect from compact binaries, results in a change in the relative velocity between freely falling test masses, and has received little attention to date. 
There are other velocity memory effects which have received some attention in the literature ~\cite{Zhang:2017rno, Zhang:2017geq, Grishchuk:1989qa, Chakraborty:2019yxn, Shore:2018kmt, Harte_2015, Lasenby:2019gmi, PhysRevD.101.024006, universe4070074, Cvetkovic:2019hep, Flanagan:2019ezo, Braginsky:1986ia}.  
These velocity memory effects can arise in several distinct situations, and their order in the gravitational-wave strain is dependent on factors such as test-mass initial conditions and the support of the strain profile, as detailed in the contents of Table \ref{tab:1}.
The results presented in this paper are consistent with discussion on the first-order velocity memory effect in \cite{Grishchuk:1989qa, Braginsky:1986ia, Zeldovich:1974gvh, Lasenby:2019gmi}, the second-order treatment in \cite{Harte_2015}, and the numerical results of \cite{Zhang:2017rno}.

The contents of this paper are outlined as follows:
In the next section, we derive the polarizations for the non-linear memory strain far away from a CBC source. 
Next, we motivate a plane wave spacetime that is useful for calculating velocity memory effects.
We then integrate the equations of motion for a pair of test masses under the influence of a polarized gravitational wave in our chosen plane wave spacetime. 
In Sec. \ref{sec:discussion}, we discuss the lessons learned from calculating the velocity memory effect for a generalized sandwich wave pulse profile.
In Sec. \ref{sec:result}, we estimate the magnitude of the first-order velocity memory effect from a typical CBC source amenable to detection by LISA.
The Appendix is structured as follows: Sec. A contains the full derivation of the non-linear memory strain and Sec. B provides the coordinate transformations between various plane wave spacetimes.

\section{\label{sec:method}Framework}

Before we present the results of the velocity memory effect from sandwich waves in Sec. \ref{sec:discussion} and the first-order velocity memory effect from compact binary sources in Sec. \ref{sec:result}, we give a summary of the framework for calculating memory effects from general gravitational-wave strain profiles.
In Sec. \ref{subsec:IIA}, we derive the plus and cross polarizations of the non-linear memory strain in a transverse-traceless (TT) gauge, by projecting onto a surface normal to the direction of propagation at the location of the detector.
In Sec. \ref{subsec:IIB}, we discuss a family of plane wave spacetime candidates and choose an ideal basis to perform our velocity memory effect calculation. 
In Sec. \ref{subsec:IID} we integrate the equations of motion in our chosen coordinate system and discuss general characteristics of velocity memory effects.
At the end of this section, we derive useful memory formulae for calculating the velocity memory effect given a time-dependent gravitational-wave profile. 

\newcommand{\Lagr}{\mathcal{L}}
\newcommand{\tdot}{\dot{u}}
\newcommand{\xdot}{\dot{x}}
\newcommand{\ydot}{\dot{y}}
\newcommand{\zdot}{\dot{z}}

\newcommand{\integ}{\int^u _{u_0}}
\newcommand{\eval}{\Big |^u _{u_0}}
\newcommand{\hdot}{\dot{h}}
\newcommand{\hddot}{\ddot{h}}
\newcommand{\evalf}{\Big|^{t_f} _{t_0}}
\newcommand{\integff}{\int^{t_f} _{t_0}}

\subsection{\label{subsec:IIA}Non-linear Memory Strain from CBC Sources}

The complete derivation of the non-linear memory strain's polarizations is given in 
Appendix \ref{AppA}, but we include a brief summary of it below. 
We use a convenient choice of orientation between the source and detector frames as outlined in \cite{PhysRevD.45.520, Wiseman:1991ss,  Favata:2009ii, PhysRevD.50.3587}. 
Given an anisotropic gravitational-wave energy flux $dE/dtd\Omega$, we can calculate the non-linear, hereditary, memory strain at a fixed distance $R$ away from a CBC source, as experienced 
by an observer located at $\theta^* , \phi^*$ in the source's spherical coordinate system.
In a generic linearized gauge, the memory tensor is given in \cite{PhysRevD.50.3587, Talbot:2018sgr, PhysRevD.45.520,Wiseman:1991ss},
\begin{eqnarray}
h _{ij} (t, \Omega^*) = \frac{4G}{Rc^4} \int_{-\infty}^{t} dt \int _{\mathcal{S}^2} \frac{dE}{dt d\Omega'} \Bigg[\frac{n' _i n' _j}{1 - n^{'l} N_l} \Bigg] d\Omega'.
\end{eqnarray}
We use spherical angular coordinates $\theta', \phi'$, defined in the source's Cartesian coordinate frame $(x',y',z')$.
The angular integral is taken over the solid-angle $d\Omega' = \sin(\theta') d\theta' d\phi'$ on the sphere of radius $R$ centered on the source.
Above, $n' (\theta', \phi')$ is defined as the general unit radial vector directed from the source to $d\Omega'$ with components: $\Vec{n}' = \left(\cos\phi' \sin\theta', \sin\theta' \sin\phi', \cos\theta' \right)$.
The observer's angular position is encoded in $N(\theta^*, \phi^*)$, defined as the unit line-of-sight vector drawn from the source to the observer with components: $\Vec{N} = \left(\cos\phi^* \sin\theta^*, \sin\theta^* \sin\phi^*, \cos\theta^* \right)$.

To separate the time information from the angular information in the memory's integrand above, we decompose the gravitational waves on the surface of radius $R$ using a spin weighted spherical harmonic basis of spin weight minus two, with time-dependent coefficients. 
For the simple case of an equal-mass, non-spinning, binary black hole merger, the dominant contribution to the outgoing gravitational-wave energy flux can be well approximated by just the quadrupole the $(l=2,m=|2|)$ modes.
Relative to a frame at the detector and aligned with that centered on the source, we can encode the information in the gravitational-wave radiation, transverse to its direction of propagation, with a simple separation of the strain into $+,\times$ polarizations.
With this specification, and after computing the angular integrals, the cross polarization contribution vanishes. 
The plus polarization of the non-linear memory strain as a function of time and the binary's inclination angle, defined as the angle between the source's angular momentum vector and the line-of-sight to the observer, is 
\newcommand{\memconstant}{\frac{1}{192 \pi}}
\begin{align}\label{eq:h_mem}
  h^{\text{mem}}_{+} (t, \iota) ={}& \memconstant \frac{R}{c} \sin^2 \iota \Big(  17   + \cos^2 \iota  \Big) \nonumber \\
  & \times \int_{-\infty}^{t} |\dot{h}_{22} (t) |^2 dt.  \\
                  \nonumber
\end{align}
Note: the final gravitational-wave strain that arrives at the detector far away from a CBC source can be described as a linear combination of spherical harmonic strain modes, of which the $(l=2,m=0)$ mode contains the largest contribution of the non-linear memory signal \cite{Talbot:2018sgr, Khera:2020mcz, Favata:2009ii, Blanchet:2008je}.
Schematically we can think of the total detected plus polarization strain as comprised of two distinct pieces by $h_+ ^{\text{tot}} (t) \sim h_+ ^{\text{osc}} (t) + h_+ ^{\text{mem}} (t) $.
In Sec. \ref{sec:result} we use the memory contribution of Eq. \ref{eq:h_mem} to calculate the first-order velocity memory effect experienced by a pair of freely falling test masses.

In the process of obtaining Eq. \ref{eq:h_mem} we projected the memory tensor $h_{ij}$ onto a transverse-traceless basis to decompose the strain into individual polarizations $h^{\text{mem}}_+ \text{ and } h^{\text{mem}}_\times$. 
In his Living Review article \cite{Blanchet:2013haa}, Blanchet outlines a generic way of developing the transverse-traceless (TT) projection of an outgoing wave onto the surface of a sphere centered on the source. 
However, as with the polar coordinates themselves, this projection has the potential to break down when the observer lies along the coordinate axis, so a projection onto the $\mathcal{R}^2$ Cartesian plane, tangent to the sphere, may sometimes be more appropriate.
This Cartesian projection will be used to discuss the TT gauge below and also enables a more general discussion of plane-fronted waves, for which exact solutions are known in general relativity \cite{Bondi:1958aj,1926RSPSA.111...95B,Brinkmann1925,1961ZPhy..163...77K} and have been widely studied \cite{Zhang:2017geq,Chakraborty:2020uui,Harte_2015,universe4070074,PhysRevD.101.024006,Flanagan:2019ezo,Rakhmanov:2014noa,Wang:2018iig}, even in connection with gravitational-wave memory effects \cite{Khera:2020mcz, Strominger:2014pwa,Shore:2018kmt,PhysRevD.50.3587,Favata:2009ii,Pollney_2011,Talbot:2018sgr,PhysRevLett.121.071102,Islo:2019qht,Johnson:2018xly,Garfinkle:2016nhe}.  
In the next subsection, we introduce additional background on plane wave spacetimes before calculating the geodesics of particles as a function of the gravitational-wave strain profile.

\subsection{\label{subsec:IIB}Plane Wave Spacetimes}

We review the plane wave spacetime formalism in the context of calculating the velocity memory effect for a general set of gravitational-wave polarizations. 
We discuss the drawbacks associated with calculating memory effects in the commonly used transverse-traceless gauge. 
After we define a suitable set of coordinates to calculate the velocity memory effect, we use the memory effect formulae derived at the end of Sec. \ref{subsec:IID} to study the second-order effects from sandwich waves in Sec. \ref{sec:discussion} and evaluate the first-order effects for gravitational waves from CBC's in Sec. \ref{sec:result}.

We first validate the plane wave approximation, 
for the detection scenario of gravitational-wave radiation from super massive binary black hole mergers in LISA. 
Consider the scenario of a gravitational-wave detector with arm-length 
$L$, at a fixed distance $R$ away from a compact binary source, observing a gravitational wave of wavelength $\lambda_{\text{GW}}$. 
Then, to first order in the path length difference, $\Delta \sim L^2/2R $, induced by the curvature of the gravitational wave-front, we require $\Delta << \lambda_{\text{GW}}$.
Such a condition ensures that the spherical wave can be considered flat over the arm-length of the detector and the gravitational-wave phase is uniform throughout. 
With an estimate of the typical distance to a super massive binary black hole merger set to $R = 1$ giga-parsec and given a minimal gravitational wavelength of $\sim 3 \times 10^{9}$ meters for LISA, we require our detector arm-length to be $L << 10^{17}$ meters, for the plane wave approximation to be valid at the location of observation. 
For the proposed arm-length of LISA ~\cite{2017arXiv170200786A}, this condition is well satisfied.

Finding a useful plane wave spacetime  to carry out the physical predictions of general relativity is a little tricky, 
as history has shown, with the infamous plane wave spacetime dispute between \cite{EINSTEIN193743} and \cite{Bondi:1958aj}, later considered by Richard Feynman who coined this problem as the ``sticky bead problem" in the famous Chapel Hill conference of 1957 \cite{alma991033133907105251}.
Examples of plane wave spacetimes appropriate for our discussion are given in the Baldwin-Jeffery-Rosen (BJR) coordinate system~\cite{1926RSPSA.111...95B, EINSTEIN193743} (in transverse-traceless gauge), and Brinkmann ~\cite{Brinkmann1925} coordinate system (a locally Lorentz gauge ~\cite{PhysRevD.71.084003, Tarabrin:2007en}). 
The coordinate transformations between these spacetimes are summarized 
in Fig. \ref{fig:gaugeplot} of Appendix \ref{AppB}.

\subsection*{\aht{Transverse-Traceless Gauge}}


For those grounded in general relativity, the familiar linearized TT gauge is a frequent choice when studying the influence of weak, polarized gravitational waves on the proper separation of test particles.
In this gauge, the gravitational plane wave must be weak, such that $\mathcal{O}(h) << 1$.
A (local) 
line element in the TT gauge $(\bar{x}, \bar{y}, \bar{z}, \bar{t})$, with the temporal profile of the metric perturbation defined as $h(\bar{t} + \bar{z}/c)$ having two polarizations ${+,\times}$, with the wave front parallel to the $\bar{x}$-$\bar{y}$ plane and travelling in the $- \bar{z}$ direction is,
\begin{align}
  ds^2 ={}& \-c^2 d\bar{t}^2 + \left(1 + h_{+}\right) d\bar{x}^2  + \left(1 - h_+ \right) d\bar{y}^2 \nonumber\\
  &   + 2 h_{\times}  d\bar{x}  d\bar{y} +  d\bar{z}^2.
\end{align}
Although 
familiar, 
the TT gauge is not an ideal choice of coordinates 
to study the dynamics 
of test particles arising from the velocity memory effect. 
The momentum in the $\bar{x}$-$\bar{y}$ plane is explicitly conserved; if a particle begins at rest, then it remains at rest even after the wave has passed by.
This feature of the TT gauge impedes our ability to study the velocity
memory effect, for which we expect the final velocity, after the wave has passed by, to be
non-zero even for particles initially at rest.

We note in passing that the TT gauge arises as a particular example of the more general Baldwin-Jeffery-Rosen coordinates (see ~\cite{Harte_2015} for discussion) which have sometimes been used for examining exact plane wave solutions. 
However, the BJR coordinates are known to be singular, hence, we will not pursue them further here.
Instead, we consider a spacetime in which the coordinates of freely falling test masses are
time dependent, non-singular, and allow for a change in the final momentum after a gravitational wave passes by.

\subsection*{Locally Lorentz Frame}
We now consider a gauge based on 
a locally Lorentz frame $(\hat{t}, \hat{x}, \hat{y}, \hat{z})$. 
This local Lorentz frame can be thought of as a proper reference frame of an observer located at the origin, such that the metric tensor must have two defining properties: at the spatial origin (${\hat{x}} ^i = 0$), the metric tensor reduces to the flat Minkowski spacetime (local) and the first derivatives of the metric tensor also vanish (Lorentz) at the origin.
The line element in this coordinate system, under the influence of a plane gravitational wave $h(\hat{t} + \hat{z}/c)$, exactly equivalent to $h(\bar{t} + \bar{z}/c)$ (see Fig. \ref{fig:gaugeplot} in Appendix \ref{AppB} for details), is 
\begin{align}
  ds^2 ={}& \left(-1 + \frac{\Phi}{2 c^2} \right) c^2 {d \hat{t}}^2 + {d \hat{x} }^2 + {d\hat{y}}^2 + \left(1 + \frac{\Phi}{2 c^2} \right) {d \hat{z}}^2 \nonumber\\
  &   + \frac{\Phi}{c}{d\hat{t} d\hat{z}},
\end{align}
where we define the gravitational-wave potential in terms of the polarization strains as,
\begin{eqnarray}
\Phi(\hat{t} + \hat{z}/c, \hat{x},\hat{y}) \equiv \frac{1}{2}\ddot{h}_+ ( \hat{x} ^2 - \hat{y} ^2 ) + \ddot{h}_{\times} \hat{x} \hat{y},
\end{eqnarray}
where a dot represents a derivative with respect to the argument of $h$.
This locally Lorentz coordinate choice is better equipped to study the dynamical response of test masses to gravitational plane waves since the relative separation between a pair of test particles will just be the difference in their positional coordinates, as pointed out in the discussion of ~\cite{PhysRevD.71.084003}. 
In addition, the momentum in the horizontal $\hat{x} - \hat{y}$ plane is not explicitly conserved since the line element explicitly depends on the $\hat{x},\hat{y}$ coordinates.
This property has an interesting feature: if particles begin at rest before the gravitational wave, then they may not necessarily be at rest after the wave has passed by.

The $\hat{x} - \hat{y}$ equations of motion in the local Lorentz frame are,
\begin{align}
\label{xLLeom}
  \frac{d^2 \hat{x}}{d \tau^2}  ={}& \frac{(c\hat{t}'  + \hat{z}')^2}{c^2} \frac{1}{4} \left( \ddot{h} _+ \hat{x} +  \ddot{h} _{\times} \hat{y} \right),  \\
\label{yLLeom}
  \frac{d^2\hat{y} }{d \tau^2}  ={}& \frac{(c\hat{t}'  + \hat{z}')^2}{c^2}  \frac{1}{4} \left( - \ddot{h} _+ \hat{y} + \ddot{h} _{\times} \hat{x} \right),
\end{align}
where $\tau$ is an affine parameter and a prime denotes a derivative with respect to $\tau$.
Note that $c = (c\hat{t}' + \hat{z}')$ and $d^2 /d \tau^2 = (\hat{t}' + \hat{z}'/c)^2 d^2 / d\hat{t}^2$.
In the simple case of a particle confined to the $\hat{z}=0$ plane and under the influence of a linearly plus polarized gravitational wave, the acceleration in the $\hat{x}$ direction may be written as,

\begin{eqnarray}
\frac{d^2 \hat{x}}{d \hat{t}^2} = \frac{1}{4} \frac{d^2 h_+ (\hat{t})}{d \hat{t}^2} \hat{x}.
\end{eqnarray}
After integration by parts, for a general initial velocity, the velocity that test masses experience is,
\begin{eqnarray}
v_{\hat{x}} = \mathcal{O}(1) + \frac{1}{4} \dot{h}_+ \hat{x} - \frac{1}{4} h \dot{\hat{x}} + \mathcal{O}(h^2).
\end{eqnarray} 
The $\mathcal{O}(1)$ term comes from the initial velocity of a test mass. 
Here is another advantageous feature of a locally Lorentz frame: a constant non-zero gravitational-wave strain value in a wave's after-zone (such as the non-linear memory component in Eq. \ref{eq:h_mem}) contributes, at first order, to the total velocity of a test particle at a particular time, given a non-zero initial velocity.
Such a feature is important to the physical case of the test masses that comprise a gravitational-wave detector such as 
LISA, since they are never truly at rest to begin with, and are constantly influenced by extraneous sources of motion such as tidal forces, stray electrostatic fields, and thermal excitation. 
\cite{2017arXiv170200786A}. 
For the remainder of the paper, and to be consistent with the work of \cite{Zhang:2017rno, Zhang:2017geq}, we calculate the velocity memory effect in a local Lorentz frame, specifically using the Brinkmann coordinates. 

\subsection*{Brinkmann Coordinates}
We have found that in a locally Lorentz frame, the velocity memory effect is expected to naturally present itself during the wave's after-zone under general conditions. 
We introduce Brinkmann coordinates, for which the  metric is an exact solution to the full non-linear Einstein field equations, and illustrate how this set of coordinates is equivalent to a local Lorentz frame.
For simplicity, we choose our transverse spatial axes $x,y$ to be aligned with the two polarizations of the gravitational wave.
The Brinkmann coordinates ($x,y,u,v$) are  global, harmonic, and describe a gravitational plane wave where the line element is
\begin{eqnarray}
ds^2 = \delta_{ij} dx^i dx^j + 2 c^2 du dv + K_{ij}(u)x^i x^j du^2.
\label{brinkmanngauge}
\end{eqnarray}
The pulse profiles $A_+(u), A_{\times}(u)$ are free functions encoded in  $K_{ij}$ via: 
\begin{eqnarray}
K_{ij} (u) x^i x^j = \frac{1}{2} A_+(u) \left( x^2 - y^2 \right) + A_{\times}(u) xy ,
\end{eqnarray}
where $K_{ij}$, a trace-free 2x2 matrix with independent pulse polarization profiles, satisfies the two-dimensional wave equation as $\Box  (K_{ij}(u) x^i x^j) = 0$. 
The time-like variable $u$ is an affine parameter.
The trivial coordinate transformation between the Brinkmann coordinates and a locally Lorentz Cartesian frame is described by the path C in Fig. \ref{fig:gaugeplot}.
The potentials between the two coordinate systems are related by $\Phi(\hat{t} + \hat{z}/c, \hat{x}, \hat{y}) = K_{ij} (u) x^i x^j$ with $A(u) = \ddot{h}(u(\hat{t} + \hat{z}/c))$.
 
Note: although the individual momenta 
in the spatially transverse $x$- and $y$-directions are 
not conserved, 
the equation for conservation of total momentum can be used to solve for $v(u)$ along a particle's trajectory, implying   
that the $v(u)$ coordinate does not give an independent equation of motion.  
Since momentum conservation does not hold strictly in the transverse plane, we can expect 
inherent divergences in the positions of test-particles when the final positions of test masses are taken to extremes, such as $u \rightarrow \infty$.
However, this is true even in the $A (u) = 0 ~\forall u$ Newtonian case, for particles initially moving with some constant relative velocity.
We recognize the existence of these non-coordinate singularities, or physical ``divergences", when using the Brinkmann coordinates \cite{Wang:2018iig} and stress that these divergences are to be physically expected.
However, to avoid explicitly encountering such divergences in our calculations,
we  integrate our equations of motion over a real and finite domain of $u$, such as $[u_0, u_f]$, in Sec. \ref{sec:discussion}.
Here, $u_0$ will always be fixed to occur before the initial arrival of a given gravitational wave of interest, and $u_f$ (not necessarily fixed) will typically occur some time later, after the passage of the wave. 

The equations of motion in the Brinkmann coordinates effectively reduce to a classical (time-dependent spring constant) two-dimensional particle motion problem, parameterized by $u$. 
Solving the Euler-Lagrange equations yields,
\begin{align}
\label{xeom}
  \frac{d^2 x}{d u^2} ={}& \frac{1}{2} \left( A_+  x + A_{\times} y \right),  \\
\label{yeom}
  \frac{d^2 y}{d u^2} ={}&\frac{1}{2} \left( - A_+ y + A_{\times} x \right).
\end{align}
This linear coupled set of second-order ordinary differential equations are in general solved numerically given a set of initial conditions and pulse profiles. 
We can see that the polarized pulse profiles $A_+, A_\times$ in the Brinkmann coordinate system are related to the second derivatives of the polarized gravitational-wave strains in the local Lorentz frame $\ddot{h}_+, \ddot{h}_\times$ by comparing Eq. ~\ref{xeom}, Eq. ~\ref{yeom} to Eq. ~\ref{xLLeom}, Eq. ~\ref{yLLeom} respectively.

For the remainder of the paper, we consider only the subset of pulse profiles $A(u)$, such that the following limit condition is met:
\begin{eqnarray}
\lim_{\beta \rightarrow \infty} \int ^{\beta} _{- \beta} A(u) du \neq \pm \infty.
\label{limcond}
\end{eqnarray}
This limiting condition ensures that a given set of polarized pulse profiles $A_{+, \times} (u)$, will result in a final, and finite, constant velocity memory between test particles. 
This limiting condition also implies a condition on the displacement trajectories: the relative separation between a pair of test masses, in the after-zone, must diverge linearly or slower to infinity.
Note: the plus contribution of the non-linear memory  strain derived in Sec. \ref{subsec:IIA}  satisfies  this  limit  condition of Eq. \ref{limcond}, since $A(u) = \ddot{h}(u(\hat{t} + \hat{z}/c))$.

In this section, we reviewed plane wave spacetimes and introduced the Brinkmann coordinates. 
We highlighted features of the Brinkmann coordinates that make them useful for studying the velocity memory effect.
In the next section, we integrate the equations of motion in Brinkmann coordinates, and solve for the geodesics of test masses for general initial conditions.
We investigate how different initial conditions, and bounding properties of the pulse profile, give rise to velocity memory effects that arise at different orders in the perturbative expansion.


\subsection{\label{subsec:IID}Integrating the Brinkmann Equations of Motion}



\begin{table*}
    \centering
    \begin{tabular}{|c|c|c|c|}
     \hline
     $I.C.$ & B [$h$] (ex. sandwich wave) &  \cancel{B} [$h$], B[$\dot{h}$] (ex. CBC case)& \cancel{B}[$\dot{h}$], B[$\ddot{h}$] \\
     \hline
     $x(u_0)\neq 0, v_0 =0$ & $\cancel{\mathcal{O}(1)}+  \cancel{\mathcal{O}(h)} - \frac{1}{4} x(u_0) \int \dot{h}_+ ^2 du,  $ & $ \cancel{\mathcal{O}(1)} +\cancel{\mathcal{O}(h)} + \mathcal{O}(h^2),   $ & $\cancel{\mathcal{O}(1)} + \frac{1}{2} \left(\dot{h}_+ x \right)  + \mathcal{O}(h^2), $  \\ 
     \hline
     $x(u_0) \neq 0, v_0 \neq0$ & $\mathcal{O}(1)+\cancel{\mathcal{O}(h)} + \mathcal{O}(h^2)  $ & $ \mathcal{O}(1) - \frac{1}{2} \left( h_+ \dot{x} \right) + \mathcal{O}(h^2) $  & $\mathcal{O}(1) - \frac{1}{2} \left( h_+ \dot{x} \right) + \frac{1}{2} \left(\dot{h}_+ x \right)  + \mathcal{O}(h^2) $\\ 
     
     \hline
     $x(u_0) = 0, v_0 = 0$ & $\cancel{\mathcal{O}(1)} +\cancel{\mathcal{O}(h)} + \cancel{\mathcal{O}(h^2)}  $ & $\cancel{\mathcal{O}(1)} +\cancel{\mathcal{O}(h)} + \mathcal{O}(h^2)  $  & $\cancel{\mathcal{O}(1)} +\cancel{\mathcal{O}(h)} + \mathcal{O}(h^2)  $\\ 
     \hline

     $x(u_0) = 0, v_0 \neq 0$ & $\mathcal{O}(1) +\cancel{\mathcal{O}(h)} + \mathcal{O}(h^2)  $ & $\mathcal{O}(1) +\mathcal{O}(h) + \mathcal{O}(h^2)  $  & $\mathcal{O}(1) +\mathcal{O}(h) + \mathcal{O}(h^2)  $\\ 
     \hline
     
    \end{tabular}
    
  \caption{In this table, we summarize at what order terms in the velocity memory effect arise based on assumptions about the gravitational-wave strain's support and initial conditions of the test masses. These results can be seen directly in the form our  velocity equations, Eq. \ref{velmemeqx} and \ref{velmemeqyzzz}, are written. 
  The first column corresponds to the initial conditions. 
  It is implicit that there are third and higher order terms in all cases. For simplicity we include just the velocity gained in the $x$ direction and a linearly polarized wave, $h_{\times}=0$. We also include the references to the previous literature on velocity memory. We use the notation B[$f(x)$] which means the function $f(x)$ has bounded support. If the B is crossed out denoted as \cancel{B}[$f(x)$], it means this function does not have bounded support. It is also implied if a function $f(x)$ has bounded support, than so do it's derivatives. 
  The case of the Christodoulou displacement memory effect is referred to in the CBC case.
  }
  \label{tab:1}
\end{table*}

To calculate the final displacements and velocities test particles experience due to a polarized plane gravitational wave, we first solve the equations of motion over a finite domain of $u$.
We integrate the acceleration equations once, from $u_0$ to some $u<u_f$, such that 
\begin{eqnarray}
\int_{u_0} ^{u} \ddot{x} du = \frac{1}{2} \left( \int_{u_0} ^{u} \ddot{h}_{+} x du+ \int_{u_0} ^{u} \ddot{h}_{\times}y du\right),
\end{eqnarray}
\begin{eqnarray}
\int_{u_0} ^{u} \ddot{y} du = \frac{1}{2} \left( - \int_{u_0} ^{u} \ddot{h}_{+} y du+ \int_{u_0} ^{u} \ddot{h}_{\times} x du\right) ,
\end{eqnarray}
where we explicitly equate $A_{+,\times}(u) = \ddot{h}_{+,\times}(u)$ in the equations of motion, with derivatives with respect to $u$.
To carry out the integration of the terms on the right-hand side, we use integration by parts, for example, 
\begin{eqnarray}
\integ \hddot_{+} x du = \left(\hdot x \right)\eval - \left( \left( h \dot{x} \right) \eval - \integ h \ddot{x} du \right).
\end{eqnarray}
The first term on the right-hand side, that involves 
the product of the current position of the particle and the amplitude of the derivative of the gravitational wave, is the contribution to the velocity memory effect observed in previous first-order calculations \cite{Grishchuk:1989qa, Zeldovich:1974gvh, Lasenby:2019gmi}.
The second term, proportional to the product of the amplitude and the current velocity, will be the contribution to the first-order velocity memory effect from CBC's in Sec. \ref{sec:result}.
The third term on the right-hand side is 
implicitly second and higher order in the amplitude and, as we will further discuss in  Sec. ~\ref{sec:discussion}, will contain a term proportional to the energy density of the wave itself, for sandwich waves.

Substituting the right-hand side of Eq. ~\ref{xeom} and Eq. ~\ref{yeom} into the integrand above, for the 
$\ddot{x}$ and $\ddot{y}$ terms, gives the following exact solutions for the velocities test particles experience in Brinkmann coordinates:
\begin{align}
\label{velmemeqx}
  \dot{x}(u) ={}& \dot{x}(u_0) + \frac{1}{2} (\hdot_+ x)\eval - \frac{1}{2}(h_+ \dot{x}) \eval \nonumber\\
  & + \frac{1}{4} \left( \integ h_+ \hddot_+ x du + \integ h_+ \hddot_{\times} y du \right) \nonumber\\
  & + \frac{1}{2}(\hdot_{\times} y) \eval - \frac{1}{2}(h_{\times} \dot{y})\eval \nonumber\\
  &  + \frac{1}{4} \left( - \integ h_{\times} \hddot_+ y du + \integ h_{\times} \hddot_{\times} x du \right),  \\
                  \nonumber
\end{align}
\begin{align}
\label{velmemeqyzzz}
  \dot{y}(u) ={}& \dot{y}(u_0) - \frac{1}{2} (\hdot_+ y)\eval + \frac{1}{2}(h_+ \dot{y}) \eval \nonumber\\
  & - \frac{1}{4} \left( - \integ h_+ \hddot_+ y du + \integ h_+ \hddot_{\times} x du \right) \nonumber\\
  & + \frac{1}{2}(\hdot_{\times} x) \eval - \frac{1}{2}(h_{\times} \dot{x})\eval \nonumber\\
  &  + \frac{1}{4} \left( \integ h_{\times} \hddot_+ x du + \integ h_{\times} \hddot_{\times} y du \right). \\
                  \nonumber
\end{align}
By writing the equations of motion in this convenient form, it is relatively simple 
to separate out the terms that contribute to the velocity memory effect at $\mathcal{O}(h)$ and those that enter at a higher order.
A summary of which assumptions lead to explicit first-order and second-order velocity memory is given in Table \ref{tab:1}. 

The corresponding displacement curves can be calculated by integrating the above velocity equations once,
\begin{equation}
x(u) = x(u_0) + \integ \dot{x}(u) du ,
\label{dispmemeqx}
\end{equation}
\begin{equation}
y(u) = y(u_0) + \integ \dot{y}(u) du .
\label{dispmemeqy}
\end{equation}
Given a set of initial conditions $x_0, y_0, \dot{x}(u_0), \dot{y}(u_0)$ and strains $h_+ (u), h_\times (u)$, we evaluate the velocity and displacement equations above, at some finite value $u_f$. 
To clarify, we define the velocity memory effect  as the final, constant, non-zero, relative velocity gained by a pair of particles due to the presence of gravitational waves.
A 
maximal transient velocity will be imparted to each test mass 
during the passage of the wave, as the integrated velocity equations above suggest.
However, that 
is not a real memory effect and is instead just an aspect of the passing gravitational wave. 

In this section, we derived the necessary equations to calculate the velocity memory effect for arbitrary pulse profiles and initial conditions. 
In Sec. \ref{sec:result} we use the memory contribution of Eq. \ref{eq:h_mem} to calculate the velocity and displacement memory effect up to $\mathcal{O}(h)$ in the perturbative expansion.
However, prior to calculating velocity memory effects from CBC sources, we discuss the lessons learned when calculating the velocity memory effect for pulse profiles that have bounded support, such as sandwich waves, in Sec. \ref{sec:discussion}.

\section{\label{sec:discussion}Velocity Memory Effect from Sandwich Waves}

\begin{figure}[H]
    \centering
    \includegraphics[height = 7.0cm, width=8.6cm]{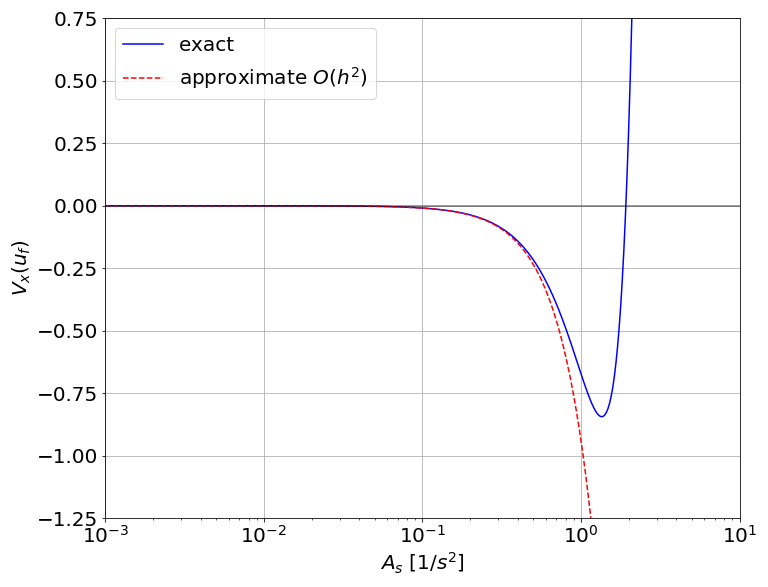}
    \caption{This plot summarizes the results of numerically integrating the Brinkmann equations of motion versus approximating the equations of motion with perturbation theory. 
    The red dotted curve (approximate) shows how much the $\mathcal{O}(h^2)$ energy density term contributes to the total velocity memory effect at a fixed amplitude of the sandwich wave. 
    The solid blue curve (exact) is calculated from numerically integrating the metric, which includes all terms. 
    We begin to see agreement between the two curves as the amplitude decreases below $A_s \sim 0.5 ~\text{sec}^{-2} $, implying that for weaker gravitational sandwich waves, the velocity memory is well approximated by just the energy density term.
    Higher order terms are responsible for the disagreement between the two curves for larger amplitudes. 
    For reference, the sandwhich wave studied in \cite{Zhang:2017rno} had an amplitude of $A_s=1 ~\text{sec}^{-2}$.
    }
    \label{fig:quadvexact}
\end{figure}

We highlight characteristics of the velocity memory effect which arises for initially stationary test masses, under the influence of a gravitational sandwich wave. 
We compare results from numerically integrating the equations of motion, and using perturbation theory to approximate the final velocity, in Fig. \ref{fig:quadvexact}.
In this section, we show that the velocity memory effect that arises in the sandwich wave case is second, and higher, order in the amplitude of the wave.

A common pulse profile to study particle orbits in Brinkmann coordinates is the sandwich wave, as examined in \cite{Zhang:2017rno} and \cite{Chakraborty:2019yxn}. 
The sandwich wave pulse profile has the general form of 
\begin{eqnarray}
A_+ (u) = \ddot{h}_{+}(A_s, \sigma;u) = \frac{d^3}{d u^3} \left(A_s e^{-\frac{1}{\sigma}u^2} \right).
\label{sandwichwave}
\end{eqnarray}
To calculate the velocity memory effect from sandwich waves, we input Eq. \ref{sandwichwave} into the right-hand side of Eq. \ref{velmemeqx} and Eq. \ref{velmemeqyzzz}, specify the initial conditions between test masses, and evaluate the integrals.

We begin our experiment by placing two test particles on the $x-y$ plane: one at the origin, that undergoes no change during and after the sandwich wave, and one at an arbitrary point $(x_0, y_0)$.
We interpret the final relative velocities and positions with respect to a reference frame centered at the particle placed on the origin.
Since we integrate over a time interval $[u_0, u_f]$, which corresponds to before and after the wave, we set $\dot{h}(u_0) = \dot{h} (u_f) = h(u_0) = h(u_f) = 0$.
We explicitly make use of the following assumptions in our analysis: (i) the particle at $(x_0, y_0)$ initially begins at rest relative to the particle at the origin and (ii) $h(u)$ has bounded support. 
We simplify our calculation by considering only a linearly polarized pulse profile so that $h_{\times} = 0$ for all $u$. 
Under these assumptions, and using the exact equations of Eq. \ref{velmemeqx} and Eq. \ref{velmemeqyzzz}, we calculate the final velocity in each direction at some finite $u_f$ after the pulse's duration, where the strain and it's derivatives effectively vanish.

The velocity memory effect for sandwich waves, with initially stationary test masses, is 
\begin{eqnarray}
\label{velx2nd}
\dot{x}(u_f) = - \frac{1}{4} x(u_0) \int_{u_0} ^{u_f} \hdot_+ ^2  du +  \mathcal{O}(h^3),
\end{eqnarray}
\begin{eqnarray}
\label{vely2nd}
\dot{y}(u_f) = - \frac{1}{4} y(u_0) \int_{u_0} ^{u_f} \hdot_+ ^2  du  + \mathcal{O}(h^3).
\end{eqnarray} 
Above, we include all the terms that arise at second order in the gravitational-wave amplitude, and it is implicit that there are third and higher order terms on the right-hand sides.
Since the sandwich wave has bounded support, there is no velocity memory effect contribution at first order in $h(u)$.
The source of the second-order term in these velocity memory equations is proportional to a quantity similar to the integral of the energy density $\dot{h}^2$ of the gravitational waves ~\cite{Harte_2015}, hence, this contribution to the overall velocity memory effect does not scale linearly with the amplitude of the plane wave.


To compare the contribution of the second-order term in Eq. \ref{velx2nd} and Eq. \ref{vely2nd}, to the total velocity memory effect, we numerically integrate our equations of motion and solve for the final velocity a sandwich wave induces in a particle that begins at rest, initially at a position of $(1,1)$. 
We then vary the amplitude $A_s$ at a fixed $\sigma = 1 ~\text{sec}^{2}$ and plot the results in Fig. \ref{fig:quadvexact}. 
As the amplitude of the sandwich wave gets smaller, the velocity memory effect is well approximated by the second-order term discussed in this section. 
As the amplitude grows larger, terms that are higher order than $\mathcal{O}(h^2)$ explain the disagreement between the two curves in Fig. \ref{fig:quadvexact}.

Integrating the velocity curves $\dot{x}(u)$ and $\dot{y}(u)$ and evaluating at some $u_f$, we arrive at an equation for the displacement memory effect in this scenario as, 
\begin{eqnarray}
x(u_f) = x(u_0) \left( 1 - \frac{1}{4}  \int_{u_0} ^{u_f} d\check{u}\int_{u_0} ^{\check{u}} \hdot_+ ^2  du \right) + \mathcal{O}(h^3) ,
\end{eqnarray}
\begin{eqnarray}
y(u_f) = y(u_0) \left( 1 - \frac{1}{4}  \int_{u_0} ^{u_f} d\check{u}\int_{u_0} ^{\check{u}} \hdot_+ ^2  du \right) + \mathcal{O}(h^3) .
\end{eqnarray}
This non-zero difference in final test particle positions due to a double integral of a second-order term was actually known nearly two decades before the results of \cite{PhysRevLett.67.1486}, in \cite{misner1973gravitation}, with the use of a Kundt metric ~\cite{1961ZPhy..163...77K}.

Thus far, we have studied the velocity memory effect case for large, and moderate, amplitudes of plane waves. 
To describe memory effects in realistic space-based gravitational-wave detectors from CBC sources, where the gravitational-wave strain scales as $1/r$, we restrict our memory effect formulae to first-order in the gravitational wave.

\section{\label{sec:result} Velocity Memory Effect  from Compact Binary Sources}
In this section, we show how the non-linear memory contribution (Eq. \ref{eq:h_mem}) from CBC sources,
induces a first-order velocity memory effect in a pair of initially non-stationary test masses, such as those that comprise LISA.
We discuss some implications of observation for the velocity memory effect in LISA.

As discussed earlier in Sec. \ref{subsec:IIA}, the gravitational-wave strain from CBC's can be thought of as a combination of oscillatory and non-oscillatory modes. 
Since the $h_{22} (t)$ oscillatory mode from CBC sources has bounded support, it 
contributes to the first-order displacement memory effect only through terms that require an integration of the oscillatory mode with respect to time.
However, this contribution is negligible since the area under $h_{22} (t)$ for a non-spinning circularized binary roughly averages to zero over sufficiently many cycles. 
For simplicity, we set $\dot{h}(t_0) =  h(t_0) = 0$ since we are concerned with changes in the relative velocity, and displacement, from $t_0$ to $t_f$.
In addition, the first-order term in the velocity memory effect, proportional to the product of the initial separation and the current value of the first derivative of the strain $\dot{h}(t_f)$, vanishes for gravitational-wave radiation from CBC's after the merger takes place in-band.
The first-order velocity and displacement memory effect equations for CBC sources are, written as a difference from their respective $\mathcal{O}(1)$ Newtonian terms $N_x , N_y = x(t_0) + \dot{x}(t_0) T, y(t_0) + \dot{y}(t_0) T$, with an initial relative velocity of $v_0 = \sqrt{\dot{x}^2 (t_0) + \dot{y}^2 (t_0)}$ and total integration time of $T = t_f - t_0$, 
\begin{eqnarray}
\label{CBCvel}
\delta v =  \frac{1}{4} v_0  h^{\text{mem}}_+ (t_f, \iota),
\end{eqnarray}
\begin{align}\label{CBCdisp}
      \delta x  \equiv x(t_f) - N_x   \sim{}&  \delta x^{C} + \frac{1}{4} \dot{x}(t_0) T  h^{\text{mem}}_+ (t_f, \iota)   \nonumber\\
  & -  \frac{1}{2} \dot{x}(t_0)  \int _{t_0} ^{t_f}  h^{\text{mem}}_+ (t, \iota)   dt ,
\end{align}
\begin{align}\label{CBCdisp2}
      \delta y  \equiv y(t_f) - N_y  \sim{}&  - \delta y^{C} - \frac{1}{4} \dot{y}(t_0) T  h^{\text{mem}}_+ (t_f, \iota)   \nonumber\\
  & +  \frac{1}{2} \dot{y}(t_0)  \int _{t_0} ^{t_f}  h^{\text{mem}}_+ (t, \iota)   dt .
\end{align}

The first-order velocity memory effect from CBC's in Eq. \ref{CBCvel} is sourced from a constant final non-zero memory strain value $h^{\text{mem}}(t_f , \iota)$ which results in a linearly diverging relative separation between freely falling test masses that comprise a detector.
This first-order velocity memory effect arises due to the non-vanishing of the integral in Eq. \ref{eq:h_mem} for a CBC system. 
This is in contrast to the velocity memory effect in Eq. \ref{velx2nd} and Eq. \ref{vely2nd}, which originated simply due to the passage of a sandwich wave. 
We denote the constant displacement memory effect in Eq. \ref{CBCdisp} and Eq. \ref{CBCdisp2}, as predicted by ~\cite{PhysRevLett.67.1486}, using $\delta x^C \text{ and }\delta y^C$, where these terms are a product of the initial separation and the final memory strain value.
The two newer contributions to the 
final displacement memory effect, are multiplied by the initial velocity in Eq. \ref{CBCdisp} and Eq. \ref{CBCdisp2}, hence, vanish for particles that are initially at rest.
Above, Eq. \ref{CBCvel} to \ref{CBCdisp2} encompass the after-zone motions of test masses, such as those that comprise future space-based gravitational-wave detectors, due to the non-linear memory contribution from compact binary sources.

We now give an order of magnitude estimate of the relative separation induced by the first-order velocity memory effect, between test masses that comprise LISA, \aht{over the detector's total observation duration.}
Consider the scenario of a pair of test masses, that comprise a single arm of the LISA detector and, moving with an initial zeroth-order relative velocity of  $\mathcal{O}(\text{m/s})$ due to orbital dynamical motion of the spacecrafts \cite{2017arXiv170200786A}.
Since the velocity memory effect induces a constant drift in separation between test masses in the wave's after-zone, it will contribute to the total path-length difference measured between the pair of masses. 
This DC ($f=0$) path-length difference is on the order of $\mathcal{O}(v_0 h^{\text{mem}} T)$ m, where $T$ is the total integrated observation time.
For a gravitational-wave memory amplitude from a super massive binary black hole merger between $\max \{ {h^{\text{mem}}} \} \sim  [10^{-17}, 10^{-22}]$ (see \cite{Islo:2019qht} for predicted rates of memory amplitudes), and an observation duration of ten years, the velocity memory effect will induce DC path-length differences, between a pair of spacecrafts, that are $\mathcal{O}(10 \text{ to } 10^{-4})$ nano-meters.
Formulating a practical method to resolve this path-length difference, from known DC displacement noise contributions in LISA, is beyond the current scope of this paper.

\quad
\section{Conclusion}

In this paper, we calculated the first-order velocity memory effect from CBC sources and found that the non-linear memory strain, which remains at a constant value after the wave has passed by, is also responsible for a constant final velocity experienced by freely falling test masses initially in motion, thereby generalizing the historic results of ~\cite{PhysRevLett.67.1486, Grishchuk:1989qa, Zeldovich:1974gvh, Braginsky:1986ia}.
We showed that the velocity memory effect from compact binary sources in LISA will induce small DC path-length differences over the detector's lifetime.
We expect these differences from the velocity memory effect, to be orders of magnitude smaller than other sources of DC displacement noise, however, we leave a detailed calculation of this comparison for future work. 
Our results suggest a novel, physical, prediction of general relativity:
freely falling test masses will diverge linearly with a constant velocity, proportional to the final value of the non-linear memory strain, in contrast
to the constant displacement of the Christodoulou memory effect.


\begin{center}
\textbf{Acknowledgements}
\end{center}
We are thankful to Guido Mueller for helpful discussion on the mechanics and measuring capabilities of LISA. 
We are grateful to Hector Chen and Nathaniel Strauss for assistance in performing calculations using mathematical computing software.
The authors gratefully acknowledge the support from the University of Florida through the National Science Foundation (NSF) grants PHY-1205512, PHY-1460803, PHY-1607323. Research reported in this publication was supported by the University of Florida Informatics Institute Fellowship Program.

\appendix
\section{Non-linear Hereditary Memory Integral}\label{AppA}
Given an anisotropic gravitational-wave energy flux, we can represent 
the non-linear memory strain at fixed distance $R$ away from a CBC source, in an arbitrary linearized gauge, 
as (given in \cite{PhysRevD.50.3587} \cite{Talbot:2018sgr} \cite{PhysRevD.45.520} \cite{Wiseman:1991ss}),

\begin{eqnarray}
h _{ij} (t, \Omega^*) = \frac{4G}{Rc^4} \int_{-\infty}^{t} dt \int _{\mathcal{S}^2} \frac{dE}{dt d\Omega'} \Bigg[\frac{n' _i n' _j}{1 - n^{'l} N_l} \Bigg] d\Omega'.
\end{eqnarray}

\begin{figure}[H]
    \centering
    \includegraphics[height = 8.0cm, width=8.0cm]{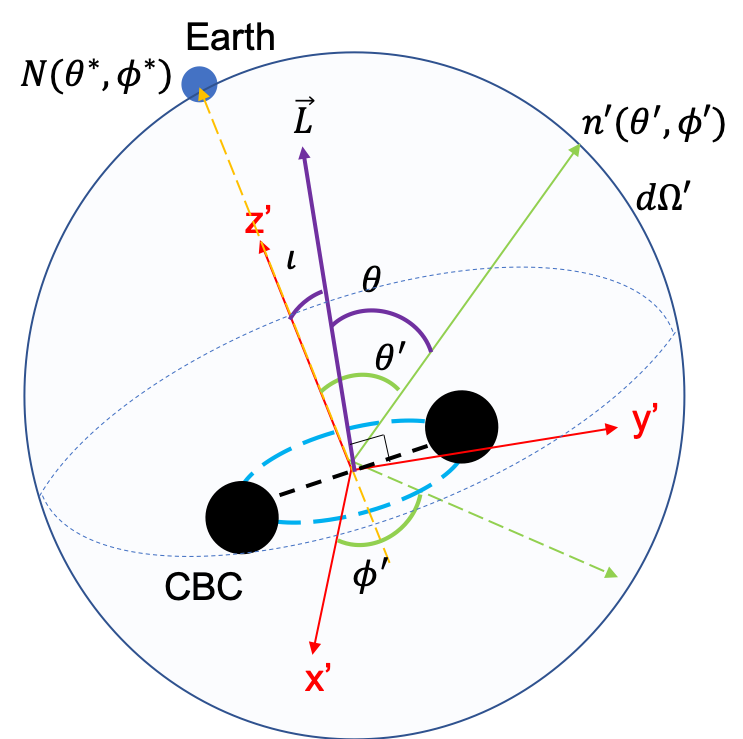}
    \caption{This diagram summarizes the orientation convention used to integrate over the solid angle in the memory equation. 
    We align the Earth along the CBC source's $z'$ axis. 
    $\Vec{L}$ is the direction of angular momentum of the CBC and is perpendicular to the plane of rotation. 
    }
    \label{fig:memdiagram}
\end{figure}

We use spherical angular coordinates $\theta', \phi'$, defined in the source's Cartesian coordinate frame $(x',y',z')$.
The angular location of the observer on the sphere $(\theta^*, \phi^*)$, centered with respect to the source, is at a distance $R$ away from the source.
The angular integral is taken over the solid-angle $d\Omega' = \sin(\theta') d\theta' d\phi'$.
Above, $n' (\theta', \phi')$ is defined as the general unit radial vector directed from the source to $d\Omega'$ with components: $\Vec{n}' = \left(\cos\phi' \sin\theta', \sin\theta' \sin\phi', \cos\theta' \right)$.
The observer's angular position is encoded in $N(\theta^*, \phi^*)$, defined as the unit line-of-sight vector drawn from the source to the observer with components: $\Vec{N} = \left(\cos\phi^* \sin\theta^*, \sin\theta^* \sin\phi^*, \cos\theta^* \right)$.
Following the convention of ~\cite{PhysRevD.50.3587} and ~\cite{PhysRevD.45.520} we set the direction from the source to the observer along the source's $z'$ axis so that the unit vector $\Vec{N} = (0,0,1)$. 
This simplifies the dot product in the denominator as $1 - n^{'l} N_l = 1 - \cos\theta'$. 
We include a diagram of the orientation convention used to derive the final memory strain in Fig. \ref{fig:memdiagram}.

For a CBC source, the gravitational-wave energy flux is defined as ~\cite{Talbot:2018sgr}, 
\begin{eqnarray}
\frac{dE}{dt d\Omega} = \frac{R^2 c^3}{16 \pi G}  |\dot{h} (t, \Omega)|^2.
\end{eqnarray}
Here, the complex valued $h(t, \Omega)$ $=h_+(t, \Omega)-ih_{\rm x}(t, \Omega)$ is the 
gravitational-wave strain, where we have used short-hand notation to describe the real part by taking the complex conjugate on the right-hand side so that the net energy flux is real and always positive. The derivative is with respect to time $t$.
Gravitational waves can be decomposed onto a spin-2 weighted spherical harmonic basis $(s=-2,l,m)$ so that we can separate the time information from the angular information, 

\begin{eqnarray}
h(t, \Omega) = h_{lm} (t) \mathcal{Y}^{lm} _{-2} (\Omega).
\end{eqnarray} 
Including multiple higher harmonic spherical modes in the memory integral , which enriches the phenomenology of the memory strain,
has been carried out 
in ~\cite{Talbot:2018sgr},
however, for our calculation, using the dominant quadrupole $(l=2,m=|2|)$ mode as the primary driver of the energy-flux is sufficient. 
In actuality, the non-linear memory part of the outgoing gravitational-wave radiation contributes to the overall energy flux, however, compared to the first-order memory signal, this memory of memory has been shown to be orders of magnitude smaller, and hence for our calculation negligible
~\cite{Talbot:2018sgr}.

The $(l=2,m=\pm 2)$ spin-weighted spherical harmonics are given by ~\cite{Garfinkle:2016nhe}, 
\begin{eqnarray}
\mathcal{Y}^{2 \pm 2} _{-2} (\theta, \phi) = \frac{1}{8} \sqrt{\frac{5}{\pi}} (1 \pm \cos \theta)^2 e^{\pm 2 i \phi}.
\end{eqnarray}
Excluding higher order mode contribution, the approximated gravitational-wave strain from the dominant quadrupole strain modes is, 
\begin{eqnarray}
h(t, \theta, \phi) \sim h_{22} (t) \mathcal{Y}^{22} _{-2} (\theta, \phi) + h_{2 -2}(t) \mathcal{Y}^{2 -2} _{-2} (\theta, \phi).
\end{eqnarray}
To simplify our calculation, we consider non-spinning equal-mass binaries, which allows us to relate the $(l=2, m=\pm 2)$ modes via the complex conjugate $h_{2 -2}  = h^* _{22} $ and justifies the exclusion of higher order modes in the energy flux.
This gives us the following expression for the magnitude of the derivative of the gravitational-wave strain, 
\begin{eqnarray}
|\dot{h}|^2 \sim \left(\dot{h}_{22} \mathcal{Y}^{22} _{-2} + \dot{h}^* _{22} \mathcal{Y}^{2 -2} _{-2}   \right) \cdot \left( \dot{h}_{22} \mathcal{Y}^{22} _{-2} + \dot{h}^* _{22} \mathcal{Y}^{2 -2} _{-2} \right)^* .
\end{eqnarray}
Following the discussion of ~\cite{Garfinkle:2016nhe} and the considerations of ~\cite{PhysRev.131.435}, after only including the terms that are non-negligible averaged over one orbit, the magnitude of the derivative of the strain becomes, 
\begin{eqnarray}
|\dot{h} (t, \theta)|^2  \sim  2 \left(\frac{5}{64 \pi}\right) \left(\cos^4 \theta + 6 \cos^2 \theta + 1 \right) |\dot{h} _{22} (t)|^2.
\end{eqnarray}

Note, the $\theta$ used in the above equation should not be confused with $\theta'$, as $\theta$ is the angle that the primary direction of emission $\Vec{n}' (\theta', \phi')$ makes with the binary's rotation axis or angular momentum vector. 
Following the orientation convention of ~\cite{PhysRevD.50.3587}, we can simplify the angular integral by orienting the source's angular momentum vector $\Vec{L}$ to lie in the $x'-z'$ plane. 
Then, we can define the inclination angle $\iota$ as the angle between $\Vec{L}$ and the direction to the observer (which we've chosen to be in the $z'$ direction).
This allows us to write the following equation relating the angle $\theta$ to $\iota, \theta', \phi'$ as outlined in \cite{PhysRevD.50.3587},
\begin{eqnarray}
\cos \theta = \sin \iota \sin \theta' \cos \phi' + \cos \iota \cos \theta'.
\end{eqnarray}
Note: we will concern ourselves only with circularized binaries.  While the energy flux would normally depend on the eccentricity, we have implicitly set that to zero.

The memory strain as a function of time and the inclination angle now becomes,
\begin{align}\label{a9}
  h _{ij} (t, \iota) ={}& \frac{5}{128 \pi^2} \frac{R}{c} \int_{-\infty}^{t} |\dot{h}_{22} (t) |^2 dt  \nonumber\\
  & \times \int _{\mathcal{S}^2} \left(1 + 6\cos^2 \theta + \cos^4 \theta \right) \Big[A' _{ij} \Big] d\Omega', \\
                  \nonumber
\end{align}
where we define the tensor, 
\begin{eqnarray}
A' _{ij} = \frac{n' _i n' _j}{1 - \cos \theta'}.
\end{eqnarray}

Next, we calculate the plus and cross polarizations of the memory strain near our detector.
Before carrying out the respective angular integrals, the angular integrand is simplified by TT projecting both the left and right-hand sides of the memory strain Eq. \ref{a9} above.
We choose a set of coordinate axes, defined at the location of our detector, and orient them so that the line of sight vector $\Vec{N}$ points along the $z'$ axis.
The TT projection on the right-hand side acts solely on the tensor $A' _{ij}$ defined above.
Using the projection operator $P_{ij} = \delta_{ij} - N_i N_j$ we can project an arbitrary symmetric tensor $B_{ij}$ into the transverse and longitudinal directions at the observer location by,
\begin{eqnarray}
B^{T} _{jk} = \frac{1}{2} P_{jk} P^{ml} B_{lm},
\end{eqnarray}
\begin{eqnarray}
B^{L} _{jk} = B_{jk} - P_{j} ^{l} P^{m} _ {k} B_{lm}.
\end{eqnarray}
Then with these components, the transverse-traceless part is, 
\begin{eqnarray}
B^{TT} _{jk} = P_{j} ^{l} P^ {m} _ {k} B_{lm} - \frac{1}{2} P_{jk} P^{ml} B_{lm}.
\end{eqnarray}

\begin{figure*}
  \centering
   \includegraphics[width=\textwidth,height=10.5cm]{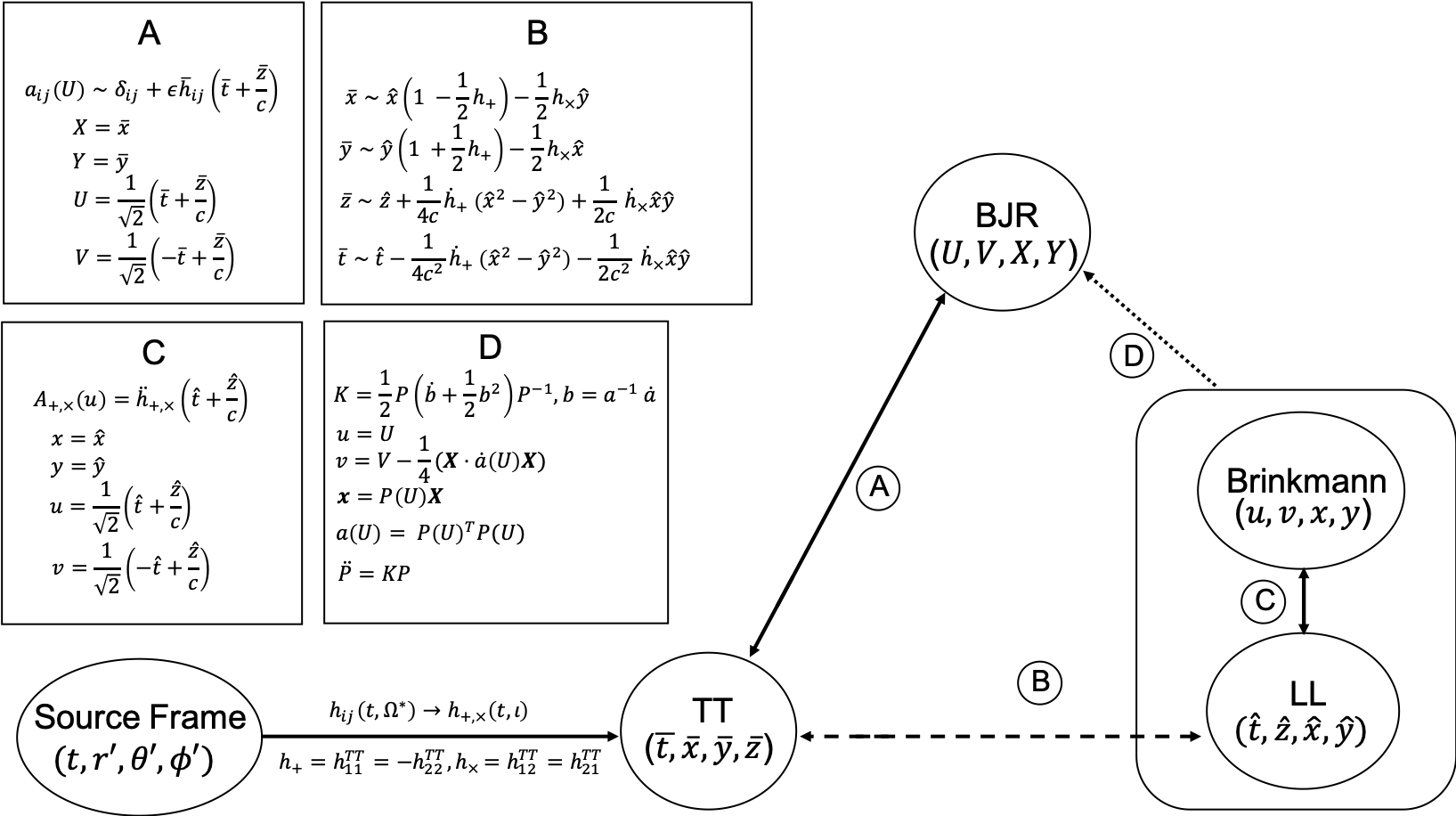}
  \caption{
  A schematic diagram showing the coordinate transformations between the CBC's source frame and the following set of plane wave spacetimes: Brinkmann, BJR, Local Lorentz (LL), and transverse traceless (TT).
  The explicit coordinate transformations are given in the four boxes to the upper left. 
  Dashed lines correspond to approximate transformations, solid lines correspond to exact transformations, and the dotted line corresponds to a transformation which can be characterized exactly, but in practice, cannot be implemented exactly.
  This diagram helps visualize the rich coordinate relations between  plane wave spacetimes.
  }
  \label{fig:gaugeplot}
\end{figure*}

We now consider the  case of a plane (or plane-projected) wave travelling in the $+z'$ direction so that all components of the projection operator vanish except $P_{11} = P_{22} = 1$. 
With this convention of the polarization basis, all the longitudinal components vanish so that the plane wave is purely transverse and traceless with polarization components, 
\begin{eqnarray}
h_+ \equiv  h^{TT} _{11} = - h^{TT} _{22},
\end{eqnarray}
\begin{eqnarray}
h_{\times} \equiv  h^{TT} _{12} = h^{TT} _{21}.
\end{eqnarray}
Hence, to calculate the plus and cross contributions of the memory strain we need to evaluate the following,
\begin{align}
  h_{+} = h^{TT} _{11} (t, \iota) ={}& \frac{5}{128 \pi^2} \frac{R}{c} \int_{-\infty}^{t} |\dot{h}_{22} (t) |^2 dt \nonumber\\
  & \times \int _{\mathcal{S}^2} \left(1 + 6\cos^2 \theta + \cos^4 \theta  \right) \Big[A' _{11} \Big] ^{TT} d\Omega', \\
                  \nonumber
\end{align}
\begin{align}
  h_{\times} = h^{TT} _{12} (t, \iota) ={}& \frac{5}{128 \pi^2} \frac{R}{c} \int_{-\infty}^{t} |\dot{h}_{22} (t) |^2 dt  \nonumber\\
  & \times \int _{\mathcal{S}^2} \left(1 + 6\cos^2 \theta + \cos^4 \theta \right) \Big[A' _{12} \Big] ^{TT} d\Omega'. \\
                  \nonumber
\end{align}
Using the TT projection operators, we can evaluate, 

\begin{eqnarray}
A^{' TT} _{11} = \frac{1}{2} \left(A_{11} - A_{22} \right) = \frac{1}{2} \left( 1 + \cos \theta' \right)  \cos 2 \phi' ,
\end{eqnarray}
\begin{eqnarray}
A^{' TT} _{12} = A_{12} = \frac{1}{2} \left(1 + \cos \theta' \right) \sin 2 \phi' .
\end{eqnarray}
Note, $A^{' TT} _{22} = - A^{' TT} _{11} = $ and $A^{' TT} _{12} = A^{' TT} _{21}$ since $A' _{ij} = A' _{ji}$.
Now we can compute the angular integrals:
\begin{align}
  F_{+} (\iota) ={}& \frac{1}{2} \int _{\mathcal{S}^2} \left(1 + 6\cos^2 \theta + \cos^4 \theta \right)  \left( 1 + \cos \theta' \right)  \cos 2 \phi'  d\Omega' \nonumber \\
  & = \frac{2 \pi}{15} \sin^2 \iota \Big(  17   + \cos^2 \iota  \Big), \\
                  \nonumber
\end{align}

\begin{align}
  F_{\times} (\iota) ={}&  \frac{1}{2} \int _{\mathcal{S}^2} \left(1 + 6\cos^2 \theta + \cos^4 \theta \right) \left(1 + \cos \theta' \right) \sin 2 \phi' d\Omega' \nonumber \\
  & = 0. \\
                  \nonumber
\end{align}

All together, we can calculate both the polarizations for the memory strain, 

\begin{align}
  h^{\text{mem}}_{+} (t, \iota) ={}& \memconstant \frac{R}{c} \sin^2 \iota \Big(  17   + \cos^2 \iota  \Big) \int_{-\infty}^{t} |\dot{h}_{22} (t) |^2 dt,  \\
                  \nonumber
\end{align}
with $h_{\times} = 0$ in this choice of polarization and basis orientation.
This result is in agreement with \cite{Favata_2010, Boersma:2020gxx, Favata:2009ii}.

\section{Coordinate Transformations Between Plane Wave Spacetimes}\label{AppB}
We provide a summary of coordinate transformations between the TT gauge, Baldwin-Jeffery-Rosen coordinates, Brinkmann coordinates, and a locally Lorentz frame in Fig. \ref{fig:gaugeplot}.

\bibliography{main.bib}
\bibliographystyle{unsrt}

\end{document}